\documentclass[english,aps,a4paper,prd,superscriptaddress,preprintnumbers,nofootinbib,preprint,floatfix,showpacs]{revtex4}

\usepackage{hyperref}%link to eqs, figs and references
\usepackage{amsmath}
\usepackage{amssymb}
\usepackage{graphicx}
\usepackage{subfig}
\usepackage{rotating}
\usepackage{color} %This is essential for placing graphics onto
\usepackage{multirow} % To use multirow command in tables
\usepackage[T1]{fontenc} 
\usepackage[utf8]{inputenc}
\usepackage{lmodern}
\makeatletter

\usepackage{booktabs}
\usepackage{mathrsfs}   %\mathscr{L}
\usepackage{slashed}     %\slashed{p}
\usepackage{url}
\usepackage{multirow}
\usepackage{units}
\usepackage{ wasysym }
\def\lfv{lepton flavour violation }
\def\lnv{lepton number violation }

\usepackage{float}
\newcommand{\bea}{\begin{eqnarray}}
\newcommand{\eea}{\end{eqnarray}}

\def\SM{$\mathrm{SU(3)_c \otimes SU(2)_L \otimes U(1)_Y}$ }
\def\ba{$$\begin{array}}
\def\ea{\end{array}$$}
\def\bra{$\begin{array}}
 \def\era{\end{array}$}

\newcommand{\g}{\,\mbox{GeV}}

\makeatother

\usepackage{babel}

\newcommand{\mutoeg}{\mu\to e\gamma}

\newcommand {\be} {\begin{equation}}
\newcommand {\ee} {\end{equation}}

\def\vev#1{\left\langle #1\right\rangle}

\newcommand{\nn}{\nonumber}

\newcommand{\AddrAHEP}{
  {\it AHEP Group, Instituto de F\'{\i}sica Corpuscular --
    C.S.I.C./Universitat de Val{\`e}ncia \\
    Edificio de Institutos de Paterna,
 C/Catedratico Jos\'e Beltr\'an, 2 E-46980 Paterna (Val\`{e}ncia) - SPAIN}}

\newcommand{\AddrLisb}{%
 Departamento de F\'\i sica and CFTP, Instituto Superior T\'ecnico\\
 Universidade de Lisboa, 
          Av. Rovisco Pais 1, 1049-001 Lisboa, Portugal }

\def\gsim{\raise0.3ex\hbox{$\;>$\kern-0.75em\raise-1.1ex\hbox{$\sim\;$}}}
\def\lsim{\raise0.3ex\hbox{$\;<$\kern-0.75em\raise-1.1ex\hbox{$\sim\;$}}}

\begin{document}

\preprint{CFTP/15-02 and IFIC/15-08}  

\title{Neutrino mass and invisible Higgs decays at the LHC }
\author{Cesar Bonilla} \email{cesar.bonilla@ific.uv.es}
\affiliation{\AddrAHEP}
\author{Jorge C. Rom\~ao}\email{jorge.romao@tecnico.ulisboa.pt}
\affiliation{\AddrLisb} 
\author{Jos\'e W. F. Valle} \email{valle@ific.uv.es}
\affiliation{\AddrAHEP}

\pacs{14.60.Pq 12.60.Fr 14.60.St }
\begin{abstract}

  The discovery of the Higgs boson suggests that also neutrinos get
  their mass from spontaneous symmetry breaking. In the simplest
  ungauged lepton number scheme, the Standard Model (SM) Higgs has now
  two other partners: a massive CP-even scalar, as well as the
  massless Nambu-Goldstone boson, called majoron.
  For weak-scale breaking of lepton number the invisible decays of the
  CP-even Higgs bosons to the majoron lead to potentially copious
  sources of events with large missing energy.
  Using LHC results we study how the constraints on invisible decays
  of the Higgs boson restrict the relevant parameters, substantially
  extending those previously derived from LEP and potentially shedding
  light on the scale of spontaneous lepton number violation.

\end{abstract}

\maketitle

\section{Introduction}

The recently discovered Standard Model (SM) Higgs boson is most likely
the first of a family.
Indeed, after the historic Higgs discovery by the LHC
experiments~\cite{Aad:2012tfa,Chatrchyan:2012ufa} it is more than ever
natural to imagine that the BEH
mechanism~\cite{Englert:1964et,higgs:1964ia,higgs:1966ev} is also the
one responsible for generating all masses in particle physics,
including those of neutrinos~\cite{Valle:2015pba}
Extra Higgs scalars are also expected in order to account for the
existing cosmological puzzles, such as dark matter and inflation, as
well as to realize natural schemes of symmetry breaking, such as those
based on supersymmetry.

Here we focus on neutrino masses. These are expected to arise from the
exchange of some heavy messenger states which, depending on the
underlying mechanism, need not be too
heavy~\cite{Boucenna:2014zba,joshipura:1992hp}.
If lepton number is broken through a \SM singlet vacuum expectation
value~\cite{Chikashige:1980ui,Schechter:1981cv} there is a physical
pseudoscalar Nambu-Goldstone boson — the majoron. 
All majoron couplings to SM particles are very small except, perhaps,
those with the Higgs boson.
As a result the CP even Higgs scalars have sizeable “invisible”
decays, for
example,~\cite{Joshipura:1992ua,romao:1992zx,joshipura:1992hp}
\begin{equation}
  \label{eq:hjj}
h \to JJ,   
\end{equation}
where $J\equiv \sqrt{2}\, \text{Im}\,\sigma$ denotes the associated
pseudoscalar Goldstone boson — the majoron. The coexistence of such
novel decays with the SM decay modes affects the Higgs mass bounds
obtained~\cite{DeCampos:1994rw,deCampos:1996bg,Abdallah:2004wy,abdallah:2003ry},
as well as provide new clues to the ongoing Higgs boson searches at
the LHC.

%%%%%%%%%%%%%%%%%%%% aqui %%%%%%%%%%%

Current LHC data suggest that the new particle discovered with a mass
$m=125$ GeV~\cite{Aad:2012tfa,Chatrchyan:2012ufa} is indeed the
long-awaited for Standard Model (SM) Higgs boson ($m_H=m$).
This places restrictions on the extended Higgs sector providing
neutrino masses, which we now analyse. We find that, despite the data
accumulated so far at the LHC, the possibility of having an invisibly
decaying  Higgs boson is not too tightly constrained.
Experimental searches have been mainly motivated by dark matter models
where the Higgs might decay into the dark matter candidate, say
$\chi$, if its mass is $m_\chi<\frac{m_H}{2}$, such as supersymmetric
models with R-parity conservation.
However, invisible Higgs boson decays appear most naturally in
low-scale models of neutrino mass generation. In these models neutrino
masses arise from the spontaneous breaking of an additional $U(1)$
global symmetry associated to lepton number in the \SM theory. This
symmetry is broken when a lepton-number-carrying scalar singlet
$\sigma$ gets a non-zero vacuum expectation value (vev), i.e.
$\langle\sigma\rangle=v_{1}$.

There are many genuine low-scale neutrino mass
scenarios of this type~\cite{Boucenna:2014zba}, such as
inverse~\cite{mohapatra:1986bd,gonzalezgarcia:1989rw} or
linear~\cite{akhmedov:1995ip,akhmedov:1995vm,Malinsky:2005bi} seesaw
schemes.
For simplicity, however, one may take the simplest \SM extension of
neutrino mass generation, namely the type-I seesaw
mechanism~\cite{GellMann:1980vs,yanagida:1979,mohapatra:1980ia,Schechter:1980gr,Lazarides:1980nt}.
In this case in order to account for the small neutrino masses one
must assume very small Dirac-type Yukawa couplings.
The important consequence of spontaneous breaking of lepton number is the
appearance of a physical Goldstone
boson~\cite{Chikashige:1980ui,Schechter:1981cv}, and the decays in
Eq.~(\ref{eq:hjj}).
The scalar sector, in the simplest scenario, contains only one $SU(2)$
scalar doublet $\phi$ and a singlet $\sigma$, called 12-model
in~\cite{Schechter:1981cv}.  Hence there are three physical spin zero
states, the two massive CP-even scalars $H_1$ and $H_{2}$ and one
massless pseudo-scalar, the majoron $J$.
Assuming the ordering $m_{H_{1}}<m_{H_{2}}$ the most interesting case
is when $m_{H_{2}}=125~\text{GeV}$.
In this letter we focus on the possibility that the Higgs $H_{2}$ is
the one reported by the LHC\footnote{The latest results from LHC
  for the Higgs boson mass are $125.36 \pm 0.37$ GeV from
  ATLAS~\cite{Aad:2014aba} and $125.02 + 0.26 - 0.27\, ({\rm stat}) +
  0.14 - 0.15\, ({\rm syst})$ GeV from
  CMS~\cite{Khachatryan:2014jba}.}, i.e. $m_{H_2}=125\g$, and that in
general the CP-even scalars can decay into majorons as follows,
\begin{equation}
  \label{eq:chain}
  H_{i}\to JJ\ \
\text{and}\ \ H_{2}\to 2H_{1}\to 4J\ \ \left(\text{when}\ \
  m_{H_{1}}<\frac{m_{H_2}}{2}\right),
\end{equation}
We note that there are strong constraints on invisible decays of a
scalar with mass below $\sim 115\g$ coming from the searches carried
out by LEP~\cite{Abdallah:2004wy}.
In the next section we describe the main features of the symmetry
breaking sector of the 12-model. 
We present our results in section III, and we discuss how the main
features of this simplest model can also be present in other schemes
with additional experimental signatures in section IV. We conclude in
section V.

\section{Symmetry breaking in the 12-model}

The simplest way to model spontaneous lepton number violation
contains, in addition to the usual SM Higgs doublet $\phi$,
\begin{equation}
  \label{eq:2}
 \phi =
 \begin{bmatrix}
   \phi^0\\
   \phi^-
 \end{bmatrix}
\end{equation}
a complex lepton-number-carrying scalar singlet $\sigma$ that acquires
a non-zero vev $\vev{\sigma}$ that breaks the global $U(1)_{L}$
symmetry ~\cite{Chikashige:1980ui,Schechter:1981cv}. This scalar gives
Majorana mass to right-handed neutrinos, while $\phi$ couples to SM
fermions. This structure defines the simplest type-I seesaw scheme
with spontaneous symmetry breaking. Many other scenarios sharing the
same symmetry breaking sector can be envisaged though, for
definiteness, we assume the simplest type-I seesaw.

\subsection{The scalar potential}

The scalar potential is given
by~\cite{Joshipura:1992ua,romao:1992zx,joshipura:1992hp}
\begin{equation}
  \label{eq:1}
  V=\mu_1^2 \sigma^\dagger \sigma + \mu_2^2 \phi^\dagger \phi 
  + \lambda_1 \left(\sigma^\dagger \sigma  \right)^2 +
  \lambda_2\ \left(\phi^\dagger \phi \right)^2 +
\lambda_{12} \left(\sigma^\dagger \sigma
\right) \left( \phi^\dagger \phi \right)  
\end{equation}
The singlet $\sigma$ and the neutral component of the doublet $\phi$
acquire vacuum expectation values $v_1$ and $v_2$,
respectively. Therefore we shift the fields as
\begin{align}
  \label{eq:3}
\sigma =& \frac{v_1}{\sqrt{2}} + \frac{R_1 + i\, I_1}{\sqrt{2}}
\nonumber\\[+2mm]
\phi^0 =& \frac{v_2}{\sqrt{2}} + \frac{R_2 + i\, I_2}{\sqrt{2}}
\end{align}

Solving the minimization equations we can obtain $\mu_1^2$ and
$\mu_2^2$ as functions of the vevs, in the usual way,
\begin{align}
  \mu_1^2=& - \lambda_1 v_1^2 - \frac{1}{2} \lambda_{12} v_2^2  
\nonumber \\[+2mm]
  \mu_2^2=& - \lambda_2 v_2^2 - \frac{1}{2} \lambda_{12} v_1^2 
\end{align}

\subsection{Neutral Higgs mass matrices}

Evaluating the second derivatives of the scalar potential at the
minimum one finds, in the basis $(R_1,R_2)$ and $(I_1,I_2)$, the
CP-even and CP-odd mass matrices, $M^2_R$ and $M^2_I$ read
\begin{equation}
  \label{eq:4}
  M^2_R =
  \begin{bmatrix}
    2\lambda_1 v_1^2 & \lambda_{12} v_1 v_2 \\[+2mm]
\lambda_{12} v_1 v_2 & 2 \lambda_2 v_2^2
  \end{bmatrix}, \quad
  M^2_I =
  \begin{bmatrix}
    0 &  0 \\[+2mm]
    0 &  0 
  \end{bmatrix}
\end{equation}

As expected, the CP-odd mass matrix has two zero eigenvalues. One
corresponds to the would-be Goldstone boson which becomes the
longitudinal component of the $Z$ boson after the BEH mechanism. The
other is the physical Goldstone boson resulting from the breaking of
the global symmetry, namely the majoron $J$. Hence we have, 
\begin{equation}
  \label{eq:5}
  J= I_1, \quad G^0 = I_2\ .
\end{equation}

For the CP-even Higgs bosons we define the two mass eigenstates
$H_i$ through the rotation matrix $O_R$ as,
\begin{equation}
  \label{eq:6}
  \begin{bmatrix}
    H_1\\[+2mm]
    H_2
  \end{bmatrix}
=
O_R
\begin{bmatrix}
  R_1\\[+2mm]
R_2
\end{bmatrix}
\equiv
\begin{bmatrix}
  \cos\alpha & \sin\alpha\\[+2mm]
-\sin\alpha &\cos\alpha
\end{bmatrix}
\
\begin{bmatrix}
  R_1\\[+2mm]
R_2
\end{bmatrix}\, ,
\end{equation}
satisfying 
\begin{equation}
  \label{eq:7}
  O_R\, M^2_R\, O^T_R = \text{diag}(m^2_{H_1},m^2_{H_2})\, .
\end{equation}
One can use Eq.~(\ref{eq:7}) and Eq.~(\ref{eq:4}) in order to solve
for the parameters $\lambda_1, \lambda_2,\lambda_{12}$ in terms of the two
physical masses and the mixing angle $\alpha$. We get
\begin{align}
  \label{eq:8}
\lambda_1 =& \frac{m^2_{H_1} \cos^2\alpha+m^2_{H_2} \sin^2\alpha}{2
  v_1^2}\, , \nonumber\\
\lambda_2 =&\frac{m^2_{H_1} \sin^2\alpha + m^2_{H_2} 
\cos^2\alpha}{2 v_2^2}\, , \nonumber\\
\lambda_{12} =&
\frac{\sin\alpha \cos\alpha \
(m^2_{H_1}-m^2_{H_2})}{v_1 v_2}\, .
\end{align}

\subsection{Higgs couplings and decay widths}

The couplings of the Higgs boson to Standard Model particles get
modified according to the substitution rule
\begin{equation}
  \label{eq:9}
  h \to \sin\alpha\, H_1 + \cos\alpha\, H_2\, .
\end{equation}
In addition to these, there are two new important couplings coming
from the extended Higgs sector, namely $H_2 H_1 H_1$ and $H_i J
J$. The former is given, with our conventions\footnote{Our Higgs
  trilinear self-coupling parameters are obtained after minimizing the
  Higgs potential. In order to get the Feynman rules we have to
  multiply by $- i$.}, by
\begin{eqnarray}
g_{H_{2}H_{1}H_{1}}&=&2 v\left[3 \lambda_2\cos\alpha\sin\alpha^2
-3\lambda_1\cos\alpha^2\sin\alpha\cot\beta\right.\notag\\
&-&\left.\frac{\lambda_{12}}{8}\csc\beta(\sin(\alpha-\beta)-
3\sin(3\alpha+\beta))\right],
\end{eqnarray}
or in terms of the masses,
\begin{eqnarray}
g_{H_{2}H_{1}H_{1}}&=&\frac{1}{2v_1 v_2}(2m^2_{H_1}+m^2_{H_2})
\sin2\alpha\left(\sin\alpha v_1 - \cos\alpha v_2\right)\notag\\
&=&\frac{\tan\beta}{2v}(2m^2_{H_1}+m^2_{H_2})
\sin2\alpha\left(\cot\beta\sin\alpha - \cos\alpha\right)\notag,
\end{eqnarray}
while the couplings $H_i J J$ are given by
\begin{equation}
  \label{eq:10}
  g_{H_{i} JJ} = \frac{\tan\beta}{v}\, m^2_{H_i}\, O_R{}_{i 1}\, ,
\end{equation}
where we have defined
\begin{equation}
  \label{eq:11}
  v=v_2= \frac{2 m_W}{g},\quad \tan\beta= \frac{v_2}{v_1}\ ,
\end{equation}
are responsible for the invisible Higgs decays. The decay widths to SM
states are obtained from those of the SM with the help of the
substitution rule in Eq.~(\ref{eq:9}). On the other hand the new
widths leading to the invisible Higgs boson decays are
\begin{equation}
  \label{eq:JJ}
  H_2\to H_1 H_1 \ \ \text{and}\ \ H_i \to JJ\, ,
\end{equation}
are given by
\begin{equation}
 \label{eq:HH}
\Gamma\left(H_{2}\to H_{1}H_{1}\right)=\frac{g_{H_{2}H_{1}H_{1}}^{2}}
{32 \pi m_{H_{2}}}\left(1-\frac{4m_{H_{1}}^{2}}{m_{H_{2}}^{2}}\right)^{1/2}
\end{equation}
and
\begin{equation}
  \label{eq:13}
\Gamma(H_i \to JJ) = \frac{1}{32\pi} \frac{g^2_{H_{i} JJ}}{m_{H_i}}
\, .
\end{equation}

\section{Results}

We now discuss the constraints on invisibly decaying Higgs bosons
which follow from searches performed at LEP as well as LHC.  We focus
on the case where the Higgs $H_{2}$ is the one reported by the LHC,
i.e. $m_{H_2}=125\g$, while $m_{H_{1}}<m_{H_{2}}$. Both states may in
principle have SM-like as well as invisible decays to majorons as
given in Eq.~(\ref{eq:chain}).

\subsection{Parameter sampling procedure}

In order to cover the possibility of a Higgs boson with mass below 125
GeV, we generate points in parameter space taking $m_{H_{2}}= 125$ GeV
and $15 < m_{H_{1}} < 115$ GeV. In our simple model, the only
remaining parameters are the vev $v_1$ characterizing the spontaneous
violation of lepton number and the mixing angle $\alpha$, which we
take as,
\begin{equation}
  \label{eq:14}
  v_1 \in[500, 1500]\ \text{GeV},\quad \alpha \in [0, \pi] \, .
\end{equation}
However, as the results do not depend very much on the value of $v_1$
in that interval, we will use $v_1=1000$ GeV in most of the results
presented.

\subsection{Theoretical constraints}

The points generated must fulfill several constraints. First come the
consistency requirements for the scalar potential, namely that it must
be bounded from below and that perturbative unitarity be
respected. The unbounded from below constraint
reads~\cite{Kannike:2012pe}
\begin{equation}
  \label{eq:theory-bounds}
  \lambda_1 > 0, \quad \lambda_2 >0, \quad \lambda_{12} + 2\sqrt{\lambda_1 \lambda_2} > 0
\end{equation}
while for the unitarity we just take a simplified approach requiring
that all couplings are less than $\sqrt{4\pi}$. Certainly this can be
refined~\cite{Djouadi:2005gi}, though Eq.~(\ref{eq:theory-bounds}) is
sufficient for our current purposes.

\subsection{Constraints from invisible decay searches}

The second type of constraints comes from the LEP collider.  Searches
for invisibly decaying Higgs bosons using the LEP-II data have been
performed by the LEP collaborations. In our setup these constraints
apply to the lightest Higgs boson, $H_1$.
For the channel $e^+ e^- \to Z H \to Z b\bar{b}$ the final state is
expressed in terms of the SM $HZ$ cross section through
\begin{align}
\label{eq:defsigma}
\sigma_{hZ \to b\bar{b}Z} = &\sigma_{H Z}^{SM}\times R_{H Z}\times
BR(H \to b \bar{b}) \nn\\ 
= &\sigma_{HZ}^{SM}\times C^2_{Z(H \to  b\bar{b}) }\,, 
\end{align}
where $R_{HZ}$ is the suppression factor related to the coupling of
the Higgs boson to the gauge boson $Z$ (i.e.  $R_{hZ}^{SM}=1$ and for
the model we have $R_{H_{1}Z}=\sin^2\alpha$; note also that $ C^2_{Z(H \to
b\bar{b})}$ is independent of $m_H$). Here $BR(H \to b \bar{b})$ is
the branching ratio of the channel $H \to b \bar{b}$ which in the
model is modified with respect to the SM by the presence of the
invisible Higgs boson decay into the Goldstone boson $J$ associated to
the breaking of the global $\mathrm{U(1)_{L}}$ symmetry. 

\begin{figure}[htb]
  \centering
    \includegraphics[width=0.55\textwidth]{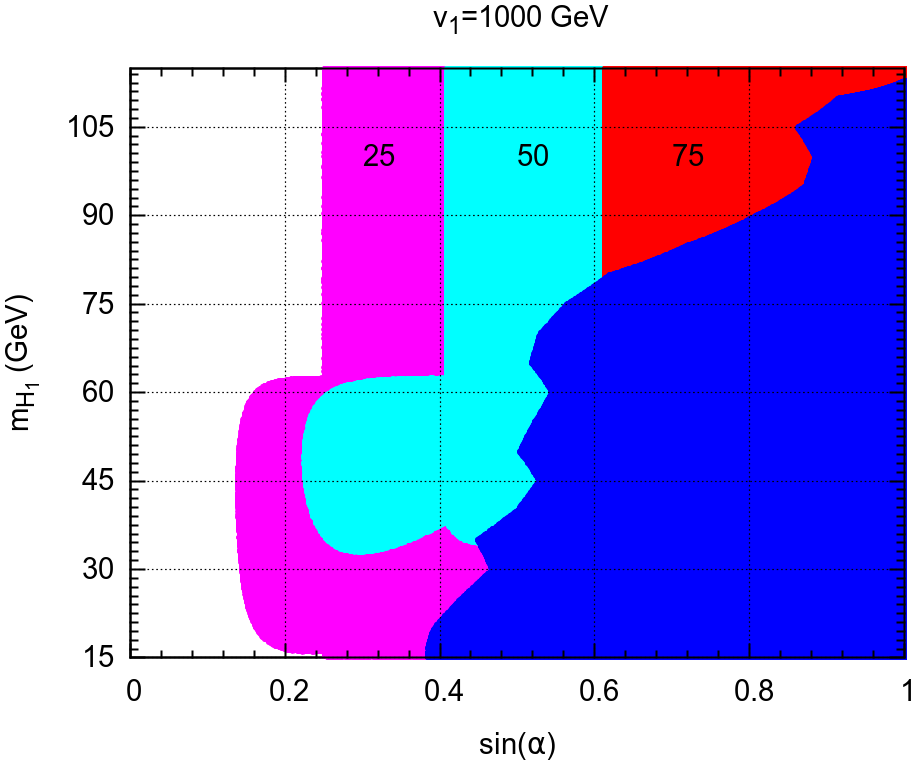}
  \caption{$m_{H_{1}}$ versus $\sin\alpha$ in the model for $v_1=1000$
  GeV. The blue region is the region excluded by LEP results. The red,
  cyan and magenta regions correspond to an invisible BR excluded at
  75\%, 50\% and 25\% respectively. }
  \label{fig:book}
\end{figure}

As illustration we consider the results from the DELPHI collaboration,
Ref.~\cite{Abdallah:2004wy}, where they give upper bounds for the
coefficients $ C^2_{Z(H \to b\bar{b})}$ corresponding to a lightest
CP-even Higgs boson mass in the range from $15$ GeV up to $100$ GeV.
From this one determines the regions of $m_{H_{1}} - \sin\alpha$ which are
currently allowed by the LEP-II searches.  The results are shown in
Fig.~\ref{fig:book}. The region excluded by the LEP results
corresponds to the blue regions in this figure.
One sees that the LEP results do not exclude much of the parameter
space for a light Higgs boson (below 115 GeV) as long as its coupling
to the $Z$ boson is reduced with respect to that of the SM. 
%%%%%
However, in this simple model, if we take into account the discovery
at the LHC of a Higgs boson at 125~GeV the parameter space is further
restricted. In fact, in this picture the heavier Higgs boson couples
to the $Z$ boson with a reduced strength $\cos\alpha$. 
The restriction on $\cos\alpha$ depends on the upper limit on the
invisible decay of the Higgs boson. Here we consider three values,
from 25\% up to 75\%, which is the current upper bound given by the
ATLAS collaboration~\cite{Aad:2014iia} for the branching ratio to
invisible particle decay modes.  This will be improved in next run of
the LHC, but current results indicate that there is still room for
such decays, as shown in Fig.~\ref{fig:book}. Note that the kink in
the plot is associated to the decay in Eq.~(\ref{eq:HH}).

\subsection{Constraints from visible decay searches}
%%\subsection{Implementing LHC constraints}

We just saw the implementation of the LHC upper limit on the invisible
decay of the Higgs boson. However we must also enforce the limits
coming from the other, well-measured, SM channels. These are normally
expressed, for a SM final state $f$, in terms of the signal strength
parameter,
\begin{equation}
  \label{eq:16}
  \mu_f =
\frac{\sigma^\textrm{NP}(pp \to h)}{\sigma^\textrm{SM}(pp \to h)}\,
\frac{\Gamma^\textrm{NP}[h \to f]}{\Gamma^\textrm{SM}[h \to f]}\,
\frac{\Gamma^\textrm{SM}[h \to \textrm{all}]}{
\Gamma^\textrm{NP}[h \to \textrm{all}]}\, ,
\end{equation}
where $\sigma$ is the cross section for Higgs production, $\Gamma[h
\to f]$ is the decay width into the final state $f$, the labels NP and
SM stand for New Physics and Standard Model respectively, and
$\Gamma[h \to\textrm{all}]$ is the total width of the Higgs boson.
These can be compared with those given by the experimental
collaborations.  We reproduce here the compilation performed in
Ref.~\cite{Fontes:2014xva} for the most recent results of the
ATLAS~\cite{Aad:2014eha} and CMS~\cite{Khachatryan:2014ira}
collaborations. One sees that the current limits, although compatible
at $1-\sigma$, still have quite large errors.

\begin{figure}[htb]
  \centering
    \includegraphics[width=0.48\textwidth]{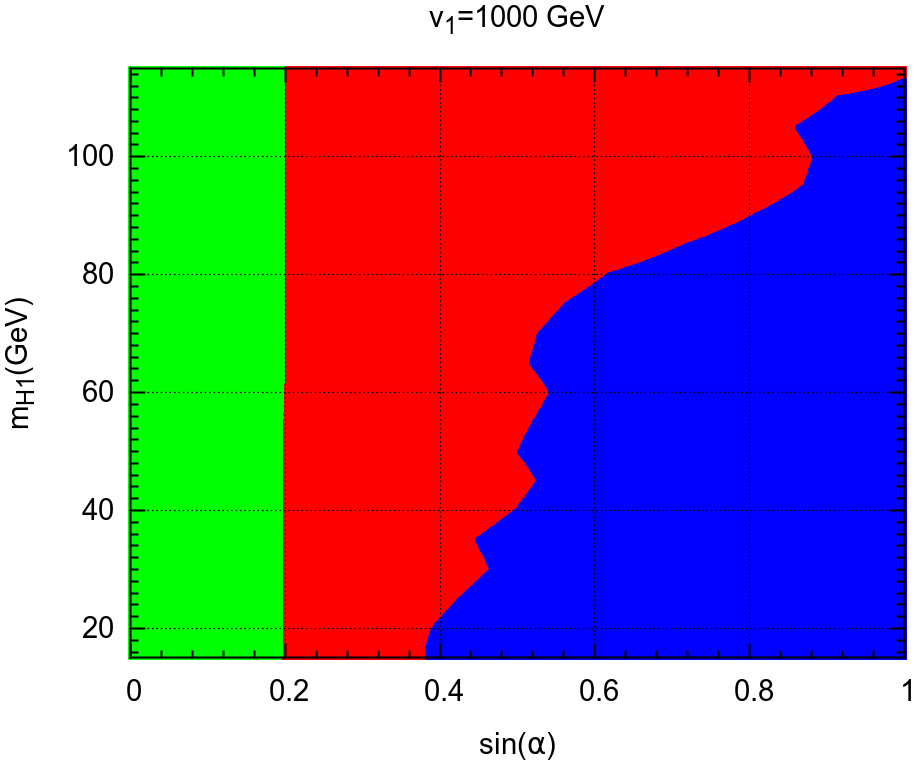}
    \includegraphics[width=0.48\textwidth]{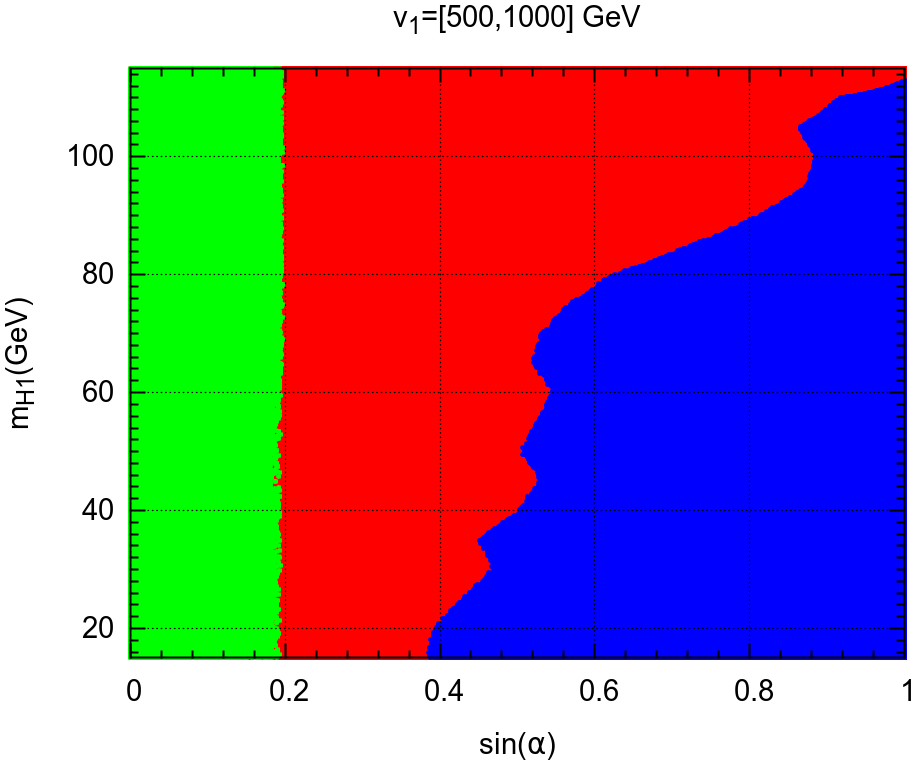}
  \caption{$m_{H_{1}}$ versus $\sin\alpha$ in the model for $v_1=1000$
  GeV (left panel) and $v_1\in [500,1000]$ GeV (right panel). The blue
  region is the region excluded by LEP results. The red corresponds to
  the points excluded by the LHC as discussed in the text and the
  points in green pass all constraints.}
  \label{fig:1}
\end{figure}

\begin{figure}[htb]
  \centering
    \includegraphics[width=0.48\textwidth]{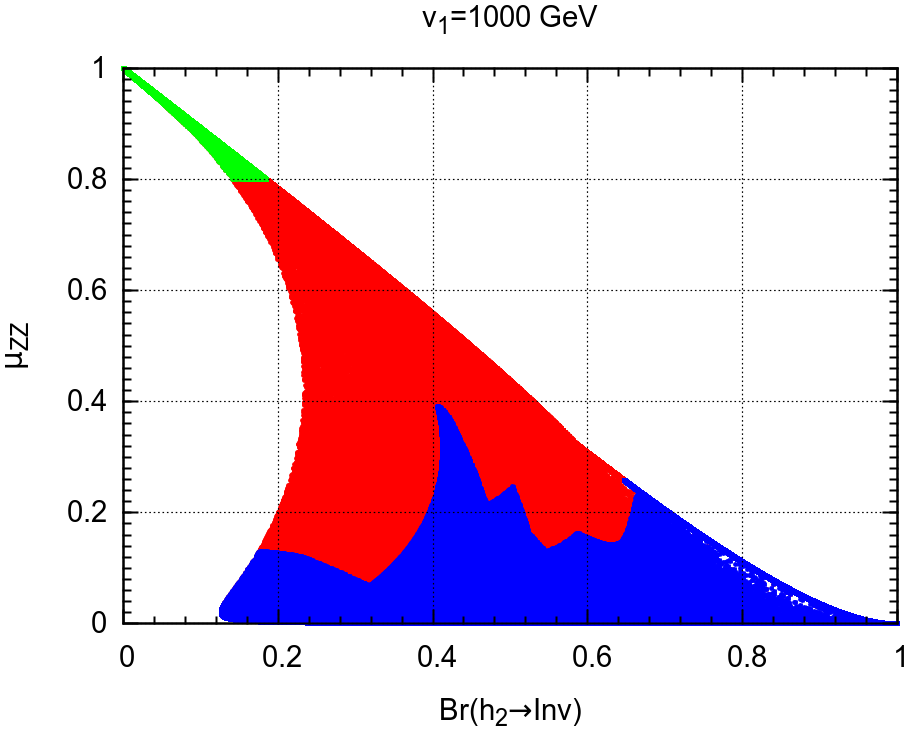}
    \includegraphics[width=0.48\textwidth]{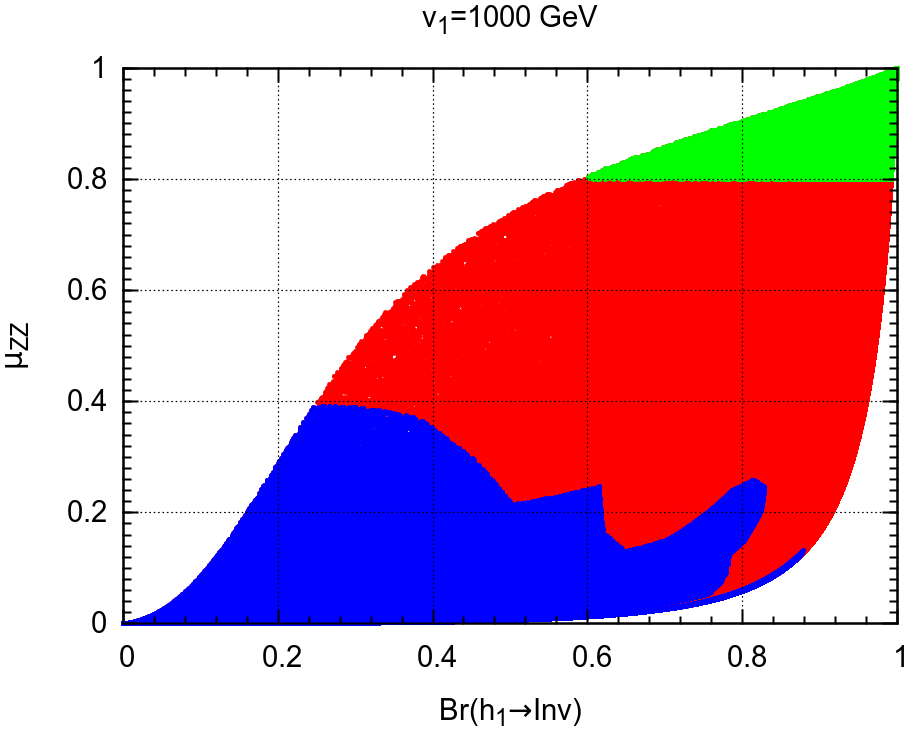}
  \caption{Left panel: $\mu_{VV}$ versus  BR($H_2\to
    \text{Inv}$). Right panel: the same for BR($H_1\to
    \text{Inv}$). The color code is as in Fig.~\ref{fig:1}.}
  \label{fig:2}
\end{figure}

\begin{figure}[htb]
  \centering
  \includegraphics[width=0.48\textwidth]{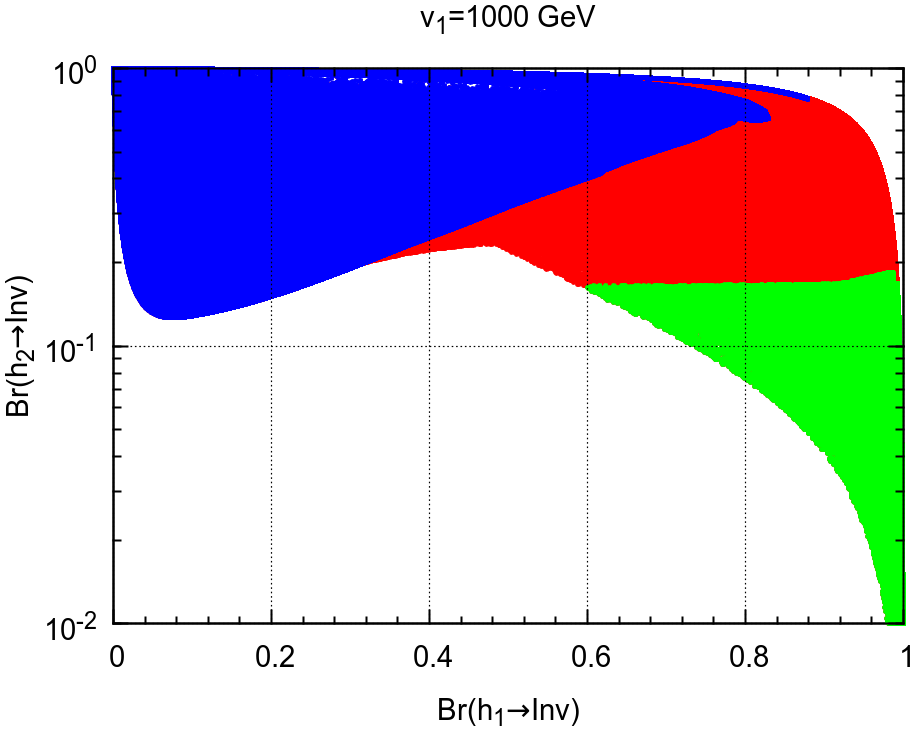}
    \includegraphics[width=0.48\textwidth]{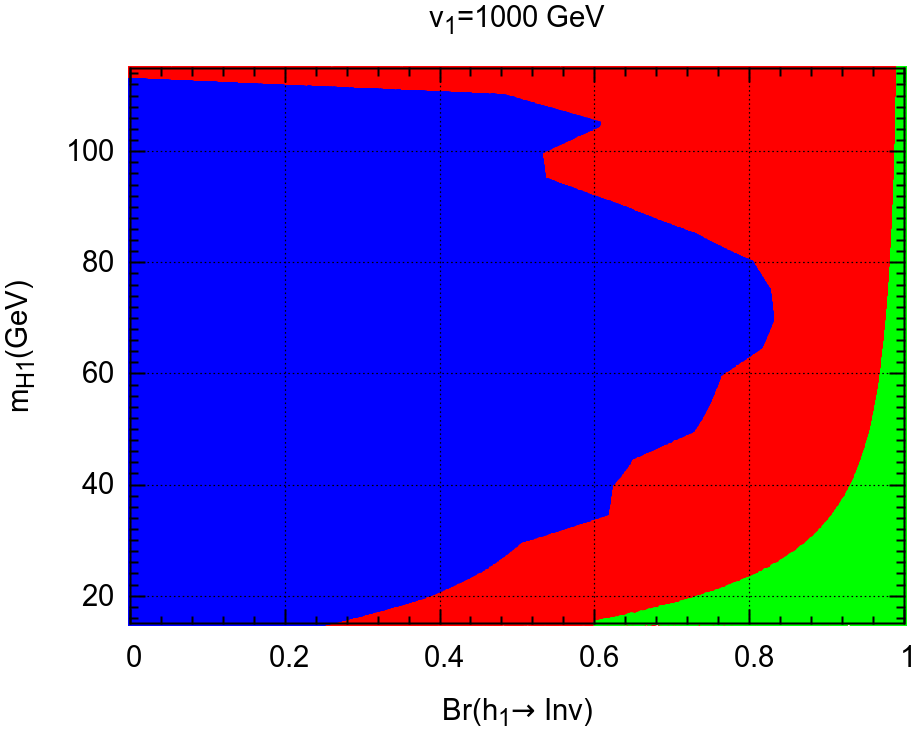}
  \caption{Left panel: BR($H_2\to \text{Inv})$ as a function of
    BR($H_1\to \text{Inv}$). Right panel: $m_{H_1}$ as a function of
    BR($H_1\to \text{Inv}$). The color code is as in Fig.~\ref{fig:1}.}
  \label{fig:3}
\end{figure}

\begin{table}[htb]
\centering
\begin{tabular}{|ccccc|}
\hline
channel & & ATLAS  & & CMS  \\
\hline
$\mu_{\gamma\gamma}$  & &
$1.17 \pm 0.27$   & &
$1.14 ^{+0.26}_{-0.23}$
\\*[2mm]
$\mu_{WW}$  & &
$1.00^{+0.32}_{-0.29}$   & &
$0.83 \pm 0.21$
\\*[2mm]
$\mu_{ZZ}$  & &
$1.44^{+0.40}_{-0.35}$   & &
$1.00 \pm 0.29$
\\*[2mm]
$\mu_{\tau^+\tau^-}$  & &
$1.4^{+0.5}_{-0.4}$   & &
$0.91 \pm 0.27$
\\*[2mm]
$\mu_{b \bar{b}}$  & &
$0.2^{+0.7}_{-0.6}$   & &
$0.93 \pm 0.49$
\\*[2mm]
\hline
\end{tabular}
\caption{\label{tab:2} Current experimental results  of
  ATLAS and CMS, taken from the compilation 
  performed in Ref.\cite{Fontes:2014xva}.}
\end{table}
Since the number of parameters is very small in our model, it suffices
to take as a constraint the limits on $\mu_{VV}$ ($V=W,Z$) in order to
illustrate the situation. Instead of taking each experiment
individually, we just note that, in a qualitative sense, the LHC
results indicate that $\mu_{VV} \sim 1$ to within 20\%, that is,
\begin{equation}
  \label{eq:17}
  0.8 \leq \mu_{VV} \leq 1.2
\end{equation}
The results are shown in Fig.~\ref{fig:1}. On the left panel we
consider $v_1=1000$ GeV while on the right panel we let it vary in the
range $v_1 \in [500,1000]$ GeV. 
As before, the blue region is the LEP exclusion region, while the red
region is excluded by the LHC limit on $\mu_{VV}$. The green
region is the region still allowed by the current LHC data. If we
compare the left panel of Fig.~\ref{fig:1} with Fig.~\ref{fig:book}
that corresponds to the same value of $v_1$, we see that the limit
imposed by $\mu_{VV}$ implies, in this model, an upper bound on
the invisible Higgs decay of around 20\%, therefore more stringent
that the one presented by the ATLAS collaboration~\cite{Aad:2014iia}. 
This is due to the fact that the
number of independent parameters is very much reduced in this model,
and the cut on $\mu_{VV}$ implies a cut on $\alpha$.
To show this, we plot in Fig.~\ref{fig:2}, $\mu_{ZZ}$ against
BR($H_i\to \text{Inv})$. The color code is as in Fig.~\ref{fig:1}.
On the left panel we see that the invisible branching ratio of the 125
GeV Higgs boson, $H_2$ in our model, could be as large as one but this
is ruled out by LEP. Furthermore the LHC limit on $\mu_{ZZ}$, reduces
the allowed space, and we obtain an upper bound on the invisible
decay, for this simple model, of around 20\% as we explained
before. The corresponding plot for the lightest Higgs boson is shown
on the right panel. We see that an invisible branching ratio of 100\%
is compatible with the LHC results for this model. The correlation
between the invisible branching ratios of the two Higgs bosons is
shown on the left panel of Fig.~\ref{fig:3} with the same convention
for the colors. Finally, on the right panel we plot $m_{H_1}$ as
function of BR($H_1\to \text{Inv}$), with the same conventions. We see
a strong anti-correlation among these panels, due to the simplicity of
the model.

\begin{figure}[htb]
  \centering
  \includegraphics[width=0.5\textwidth]{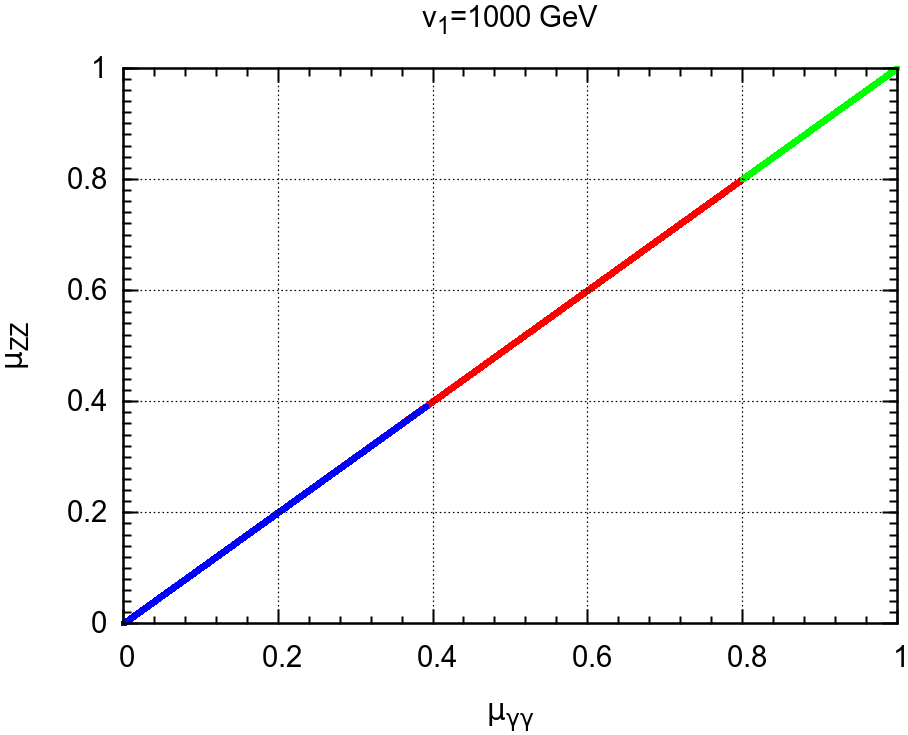}
  \caption{Correlation between $\mu_{ZZ}$ and $\mu_{\gamma\gamma}$. 
  The color code is as in Fig.~\ref{fig:1}.}
  \label{fig:4}
\end{figure}

In order to better illustrate this anti-correlation we plot in
Fig.~\ref{fig:4}, $\mu_{ZZ}$ as a function of $\mu_{\gamma\gamma}$.
The straight line reflects the fact the there is essentially only one 
parameter left, the angle $\alpha$, after we fix the two Higgs boson 
masses. We also notice that in the model, the $\mu_f$ for the channels 
where the final state $f$ exists in the SM can only be less then one. 
This results from the reduced coupling of the SM-like Higgs boson.

More general models with a richer Higgs boson sector naturally emerge,
for example, in neutrino mass schemes with more than one scalar
doublet~\cite{romao:1992zx,joshipura:1992hp, Lopez-Fernandez:1993tk}
or models with a doublet and triplet~\cite{Diaz:1998zg}. In this case,
in addition to the scalars considered here there are also charged
Higgs bosons. Similar features hold in models where the origin of
neutrino mass is supersymmetric, due to spontaneous breaking of
R-parity~\cite{Hirsch:2004rw,Hirsch:2005wd}.

\section{Discussion}

In this paper we have given a simple ``generic'' example illustrating
how the physics associated to the Higgs boson may get modified within
extensions of the minimal \SM theory with spontaneous \lnv at
low-scale~\cite{Valle:2015pba}~\footnote{High-scale seesaw models may
  lead to sizeable \lfv rates coming from supersymmetric
  contributions~\cite{antusch:2006vw}. However here we discard this
  possibility, since the sizeable invisible Higgs boson decay physics
  would be absent in that case.}.
So far we have considered the simplest scenario for spontaneous
breaking of ungauged lepton number symmetry responsible for inducing
the tiny neutrino masses. The latter involves the standard \SM
electroweak gauge structure, and hence gives rise to a physical
Goldstone boson that provides an invisible Higgs decay channel. Such
simple scheme can be implemented in a variety of ways both at the tree
level as well as within radiative
schemes~\cite{Boucenna:2014zba,joshipura:1992hp}.

Additional phenomenological signatures beyond the invisible Higgs
decay channel in Eq.~(\ref{eq:hjj}) include charged \lfv processes
such as radiative muon and tau decays, e.g. ${\mutoeg}$, $\mu \to 3e$
as well as mu-e conversion in nuclei.
The expected rates for such processes will depend on the details of
the model considered. For this reason they have not been discussed
explicitly in the present paper. For example ${\mutoeg}$ can be large
in inverse seesaw
schemes~\cite{bernabeu:1987gr,Deppisch:2004fa,Dev:2009aw}. Likewise,
mu-e conversion in nuclei is also enhanced~\cite{Deppisch:2005zm}.
Similar enhancement of \lfv processes exists for linear seesaw-type
schemes~\cite{Forero:2011pc}.
Similar features arise within radiative models of neutrino mass
generation, for example models of the
Zee-Babu-type~\cite{aristizabalsierra:2006gb}. These models include
also physical charged scalar bosons running in the neutrino mass loop,
and their scalar potential is richer than we have considered above.
Note that the charged scalar states present in such models also give a
contribution to ${H\to \gamma\gamma}$ decays.
Finally, there is a different class of charged \lfv processes
involving majoron emission, for example $\mu\to e J$. This possibility
has been considered, for example, within supersymmetric models with
spontaneous R parity
violation~\cite{santamaria:1988zm,romao:1991tp,PhysRevD.79.055023}.

\section{Conclusions}

Here we have considered the constraints implied by current data,
including the Higgs discovery, on the extended electroweak symmetry
breaking potential corresponding to the simplest neutrino mass schemes
with spontaneous breaking of lepton number.
There are two CP-even Higgs scalars that can decay to Standard Model
states as well as invisibly to the majoron, the pseudoscalar Goldstone
boson associated to lepton number violation.
If lepton number symmetry breaks at the weak scale, the invisible
modes can yield potentially large rates for missing energy events.
Using current results from LEP and ATLAS/CMS at the LHC we have
studied the constraints coming from SM searches as well as invisible
decays, showing how, despite the large data sample, there is still
room for improvement of invisible decay limits in the coming LHC
run. Within our simple framework these limits provide a probe into the
scale characterizing the violation of lepton number responsible for
neutrino mass generation.
Having set out the general strategy, other more complex symmetry
breaking sectors may be analysed in a similar way such as, for
example, those arising in models containing charged Higgs bosons.

\section{Acknowledgments}

Work supported by the Spanish grants FPA2011-22975 and Multidark
CSD2009-00064 (MINECO), and PROMETEOII/2014/084 (Generalitat
Valenciana). J.C.R. is also support in part by the Portuguese
Funda\c{c}\~ao para a Ci\^encia e Tecnologia (FCT) under contracts
PEst-OE/FIS/UI0777/2013, CERN/FP/123580/2011 and
EXPL/FIS-NUC/0460/2013.

%\bibliography{newrefs.bib,refs321.bib,merged.bib}

\end{document}